\documentclass[10pt,twocolumn,letterpaper]{article}

\usepackage{cvpr}              

\usepackage{xcolor}
\usepackage{multirow}
\usepackage{graphicx}
\usepackage{tabularx}
\usepackage{wrapfig}
\usepackage{caption}

\newcommand{\oursName}{SVHalluc}

\definecolor{ella}{rgb}{0.9764,0.447,0.447}

\definecolor{r1}{rgb}{0.0,0.2,0.6}
\definecolor{r2}{rgb}{0.0,0.4,0.2}
\definecolor{r3}{rgb}{0.6,0.0,0.0}

\usepackage{tcolorbox}
\usepackage{fancyvrb}
\tcbuselibrary{skins, breakable}
\newtcolorbox{promptbox}[1]{
    colback=gray!10,           
    colframe=black,            
    arc=3mm,                   
    boxrule=1pt,               
    left=4mm, right=4mm, top=3mm, bottom=3mm,  
    fonttitle=\bfseries,       
    title={#1},                
    breakable                  
}

\tcbuselibrary{listings}
\tcbset{
    listing style/.style={
        listing only,
        colback=white,
        colframe=black,
        boxrule=0.5pt,
        sharp corners,
        left=6pt, right=6pt, top=6pt, bottom=6pt,
        fonttitle=\bfseries,
        listing options={
            basicstyle=\ttfamily\small,
            breaklines=true,
            columns=fixed,
            keepspaces=true,
            linewidth=\textwidth
        }
    }
}

\newcommand{\delete}[1]{\textcolor{cyan}{[DELETE]}}

\definecolor{darkgreen}{RGB}{0, 150, 0}

\newcommand{\qwenthree}{\raisebox{0.1em}{\includegraphics[height=.8em,trim=0 2em 0em 0]{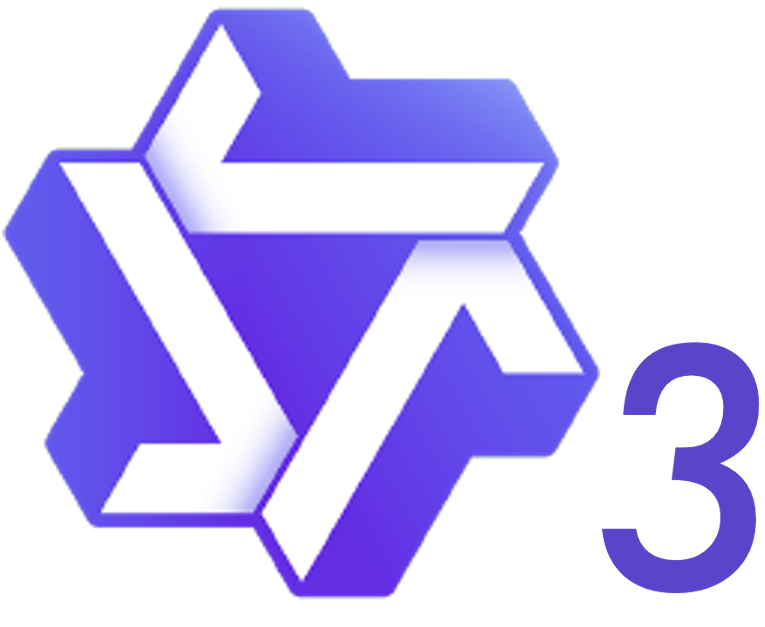}}}

\newcommand{\qwentwo}{\raisebox{0.1em}{\includegraphics[height=.8em,trim=0 2em 0em 0]{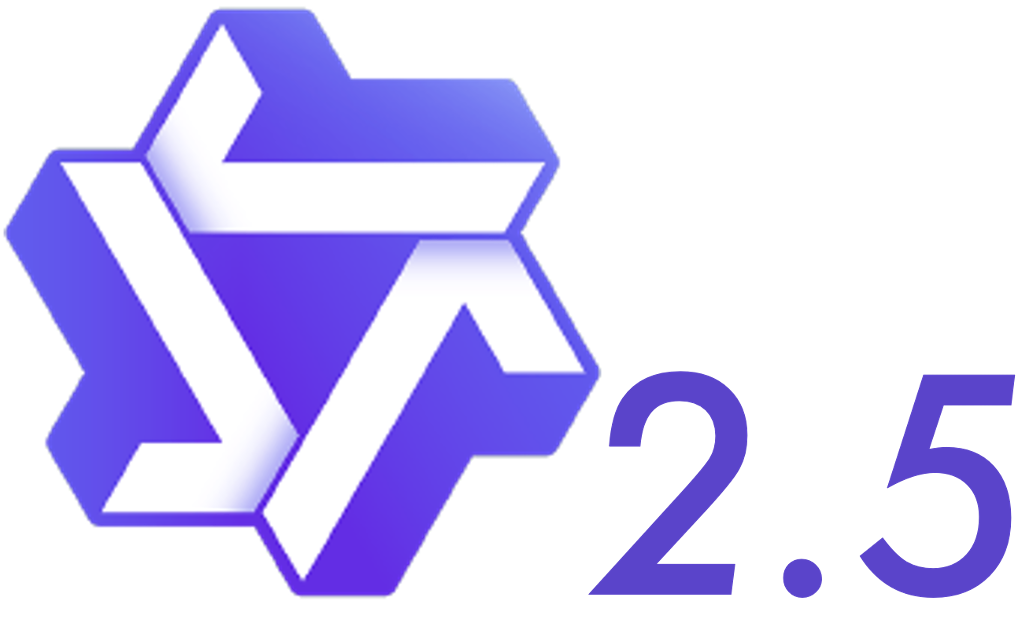}}}

\newcommand{\llama}{\raisebox{0.1em}{\includegraphics[height=.8em,trim=0 2em 0em 0]{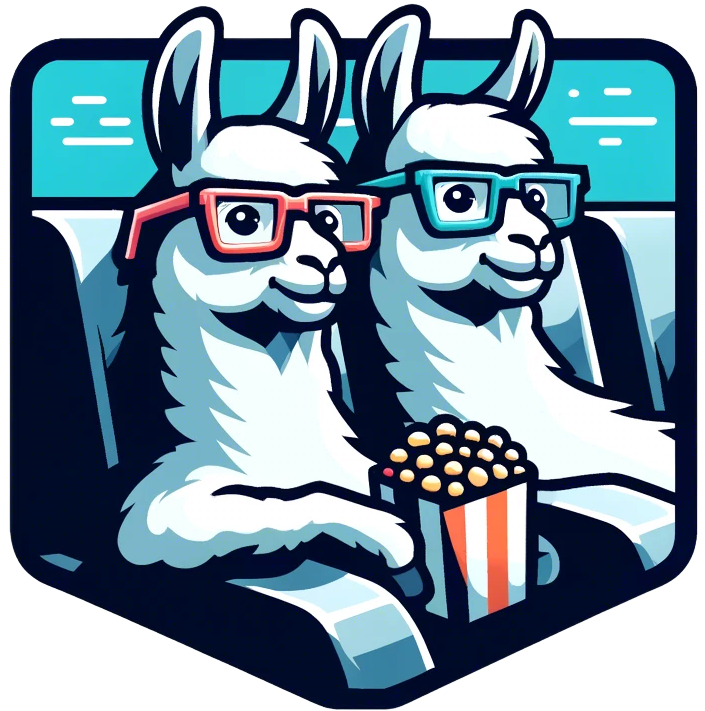}}}

\def\0{{\bf 0}}
\def\1{{\bf 1}}

\makeatletter
\newcommand{\startsuptoc}{%
  \let\orig@addcontentsline\addcontentsline
  \def\addcontentsline##1##2##3{%
    \orig@addcontentsline{suptoc}{##2}{##3}%
  }%
}

\newcommand{\tablesupcontents}{%
  \section*{Contents}
  \@starttoc{suptoc}
}
\makeatother

\definecolor{cvprblue}{rgb}{0.21,0.49,0.74}

\usepackage[pagebackref,breaklinks,colorlinks,allcolors=cvprblue]{hyperref}

\def\confName{CVPR}
\def\confYear{2026}

\title{SVHalluc: Benchmarking Speech-Vision Hallucination  \\ in Audio-Visual Large Language Models}

\author{Chenshuang Zhang \quad Kyeong Seon Kim \quad Chengxin Liu \quad Tae-Hyun Oh\\
		KAIST \\
		{\tt\small \{zcs15, taehyun.oh\}@kaist.ac.kr}
	}

\begin{document}
\maketitle
\begin{abstract}

Despite the success of audio-visual large-language models (LLMs), they can produce plausible but ungrounded outputs, termed hallucination. Existing benchmarks focus on environmental sounds (e.g., dog barking) to indicate event occurrence. In contrast, human speech carries fundamentally different, rich semantics and temporal structures, yet it remains unexplored whether current models can accurately align speech content with corresponding visual signals. In this work, we show that speech content can induce hallucinations in audio-visual LLMs. To systematically study this, we introduce SVHalluc, the first comprehensive benchmark for evaluating speech–vision hallucination in audio-visual LLMs. Our benchmark diagnoses speech–vision hallucinations from two critical and complementary aspects: semantic and temporal. Experimental results demonstrate that state-of-the-art open-source audio-visual LLMs struggle with aligning speech content with corresponding visual signals, with a near-random accuracy on multiple tasks. In contrast,  Gemini 2.5 Pro significantly outperforms the open-source models. Our analysis suggests that their failures stem from limited ability in cross-modality understanding,  despite strong performance in single-modality perception. Our work uncovers a new and fundamental limitation of current audio-visual LLMs and highlights the need for speech-grounded video comprehension. Project page: \small{\url{https://chenshuang-zhang.github.io/projects/svhalluc/.}}

\end{abstract}    
\section{Introduction}
\label{sec:intro}

Large language models (LLMs) have achieved significant advancement from text understanding~\cite{touvron2023llama,yang2025qwen3technicalreport} to multimodal reasoning~\cite{alayrac2022flamingo,Chen_2024_CVPR,zhu2024minigpt,lin-etal-2024-video,hyun-etal-2024-smile}. Audio-visual LLMs~\cite{cheng2024videollama,xu2025qwen25omnitechnicalreport,xu2025qwen3} jointly process video and audio to handle real-world multimodal scenarios. Despite impressive abilities, recent studies~\cite{sung-bin2025avhbench,radevski2025dave,leng2025the} find that audio-visual LLMs can produce plausible outputs but not grounded in the input, a failure termed as hallucination~\cite{zou2025look,jung2025avcd}. This raises significant safety concerns for the application of existing audio-visual LLMs and highlights the urgent need to evaluate how well they integrate multimodal inputs and where they may fail.

Existing audio-visual benchmarks~\cite{leng2025the,sung-bin2025avhbench,radevski2025dave} examine hallucination primarily through environmental sounds, such as dog barking or sirens. Specifically, they use environmental sounds as indicators of event occurrence~\cite{leng2025the,sung-bin2025avhbench,radevski2025dave}. 
As a result, prior benchmarks reduce audio-visual understanding to questions like ``Is the dog making sound in the audio?''~\cite{sung-bin2025avhbench} or ``What is the person doing when siren is heard?''~\cite{radevski2025dave}. Such simple question types~\cite{leng2025the,sung-bin2025avhbench,radevski2025dave} focus on limited semantic events. 
Moreover, existing benchmarks~\cite{radevski2025dave} typically use environmental sounds to indicate the current moment, without referring to a different time (e.g., past or future). 
As such, the existing benchmarks ~\cite{leng2025the,sung-bin2025avhbench,radevski2025dave} are only dedicated to such specific types of hallucination modes in audio-visual LLMs.

\begin{figure}[t!]
    \vspace{-10pt}
    \centering
    \includegraphics[width=\linewidth]{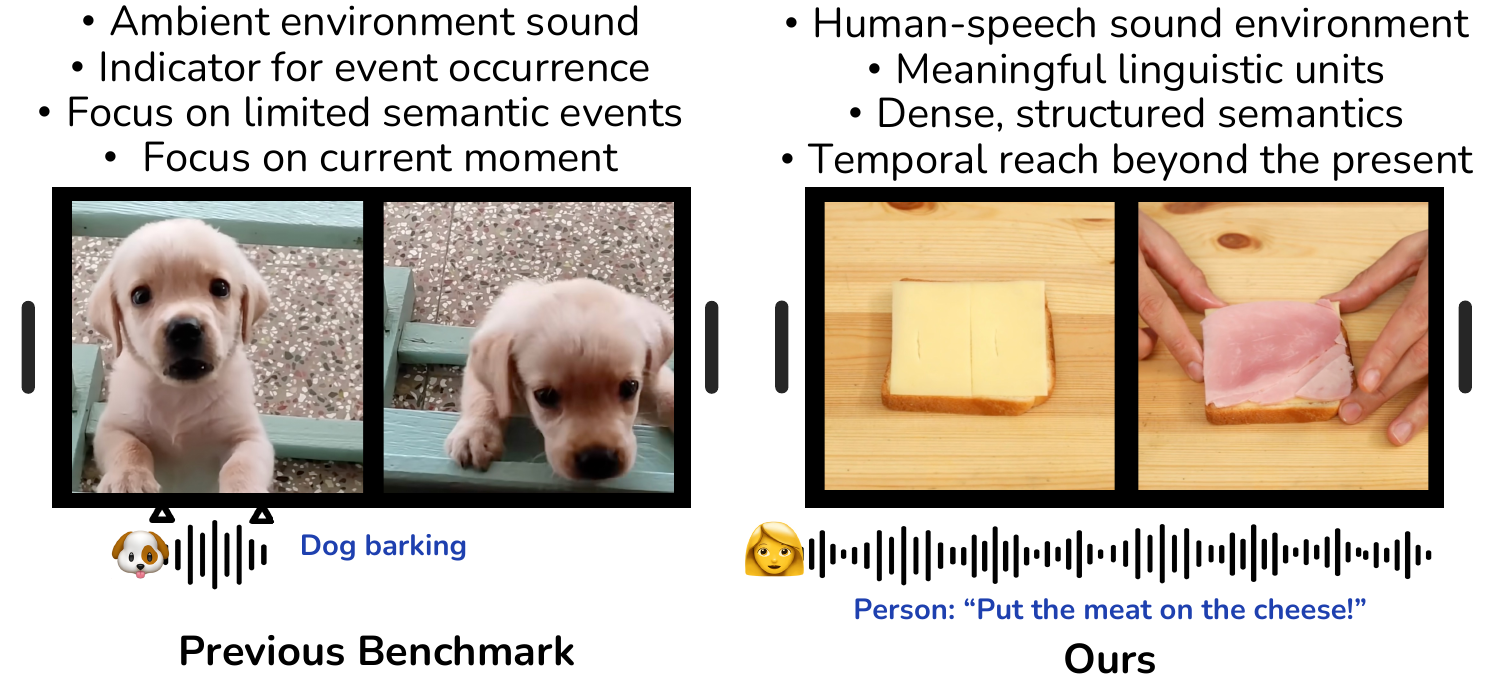}
    \vspace{-20pt}
    \caption{
        \textbf{Differences between our work and existing benchmarks.}
        Existing benchmarks evaluate audio-visual hallucination mainly by environmental sounds (e.g., dog barking), using them as indicators of event occurrence. In contrast, our benchmark investigates the speech-vision hallucination induced by the speech content. The rich semantic and temporal information conveyed in speech poses significant challenges to audio-visual understanding. Our work uncovers hallucination modes in audio-visual LLMs that previous benchmarks cannot capture.
    }
    \label{fig:Teaser}
    \vspace{-10pt}
\end{figure}

In this work, we investigate speech-vision hallucination, a setting unexplored in existing benchmarks and fundamentally different from environmental sounds. First, unlike previous benchmarks~\cite{leng2025the,sung-bin2025avhbench,radevski2025dave} that use  environmental sounds to indicate event occurrence, speech is beyond an event indicator since its content cannot be summarized as a simple ``a person is speaking.''
Second, speech conveys rich information, such as instructions or even irrelevant commentary. This introduces a complex semantic relationship between speech content and visual scene, yet is overlooked by existing benchmarks. Third, speech can describe events that happen in the past, present or future, while previous benchmarks only use environmental sounds to indicate the moment when they occur~\cite{radevski2025dave}. These properties make speech-vision understanding significantly challenging and introduce hallucination modes that previous benchmarks cannot capture.
 
\begin{figure*}[t]
    \centering
    \vspace{2pt}
    \includegraphics[width=\textwidth]{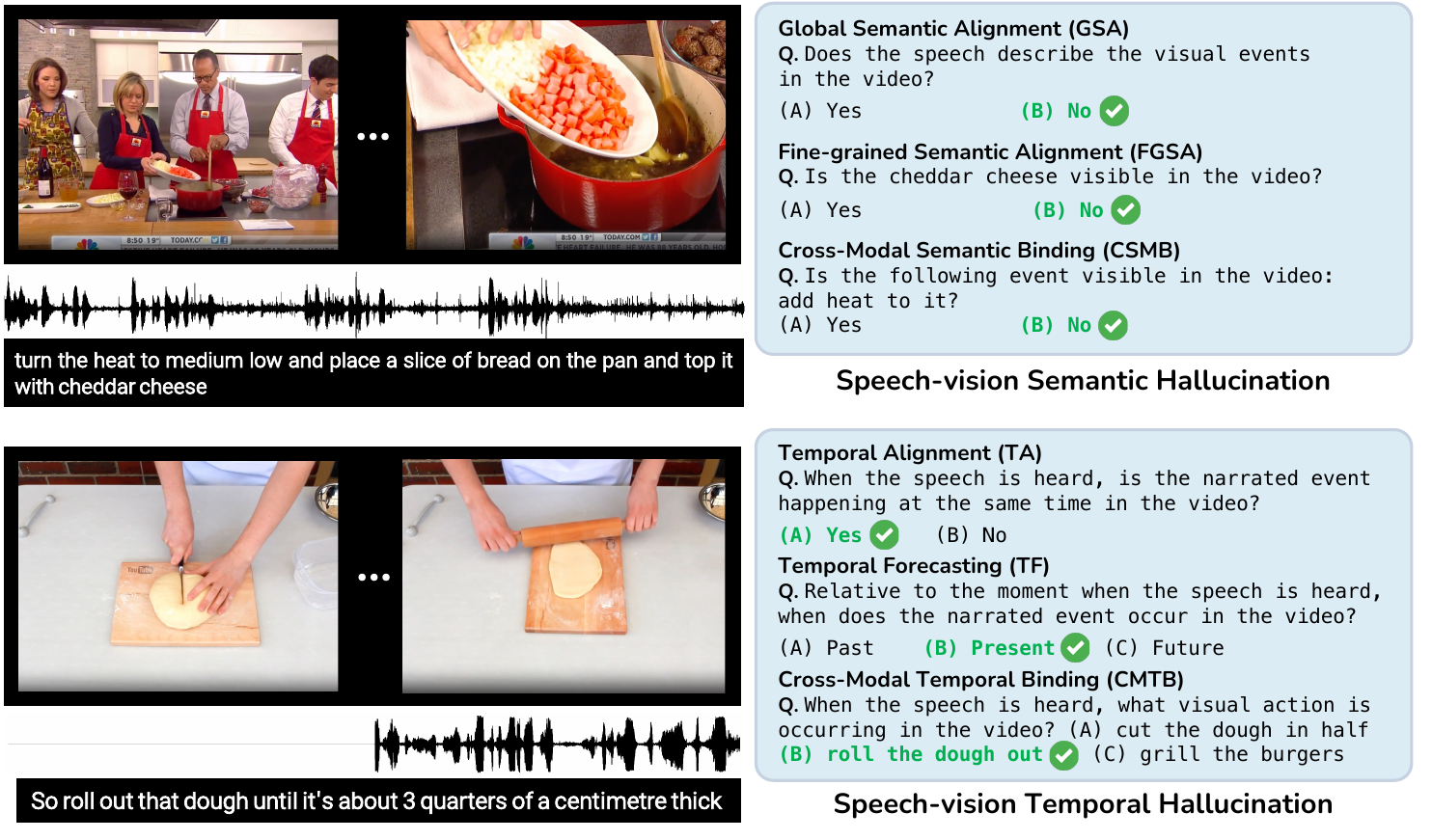}
    \vspace{-6pt}
    \caption{ 
    \textbf{Cross-modal speech-vision hallucinations in audio-visual LLMs.} Our work shows that speech content can induce hallucinations in existing audio-visual LLMs.  To systematically study this, we propose \oursName{}, the first comprehensive benchmark that investigates speech-vision hallucinations in audio-visual LLMs. Specifically, we investigate from two critical and complementary aspects: semantic and temporal. For semantic hallucination tasks, we evaluate whether models can find the semantic correspondence between speech and the visual scene, rather than hallucinating non-existing entities. For temporal hallucinations, we investigate whether models could identify the temporal relationship between narrated events and visual actions. Our work uncovers hallucination modes in audio-visual LLMs unexplored in previous benchmarks. 
    }
    \label{fig:task}
    \vspace{-4pt}
\end{figure*}

To systematically address these challenges, we introduce \oursName{}, the first comprehensive benchmark specifically designed to evaluate speech–vision hallucination in audio-visual LLMs. Specifically, \oursName{} evaluates whether audio-visual LLMs can accurately integrate speech and visual information along two critical and complementary dimensions: semantic and temporal.  In semantic hallucination tasks, we aim to evaluate whether models can correctly find the correspondence between speech content and the visual scene, avoiding hallucinating non-existing entities (e.g., objects or events). In temporal hallucination tasks, we aim to evaluate whether models could identify when narrated events visually occur compared to the moment when speech is heard.  We design three complementary coarse-to-fine tasks for each hallucination type. Our \oursName{} systematically diagnoses failure modes of current audio-visual LLMs, providing a rigorous and comprehensive evaluation of speech-vision hallucinations.

We evaluate the state-of-the-art audio-visual LLMs on our benchmark, including the open-source Qwen3-Omni~\cite{xu2025qwen3} and commercial model Gemini 2.5 Pro~\cite{Gemini2.5pro}. Experimental results show that all open-source models perform poorly on \oursName{}, often approaching random guessing. This demonstrates that existing open-source models struggle to align speech content with the corresponding visual evidence, an issue unexplored in prior studies and potentially harmful in real-world applications. The stronger performance of  Gemini 2.5 Pro~\cite{Gemini2.5pro} also indicates a huge gap between open-source and commercial models. We further conduct a comprehensive analysis of why these models hallucinate on \oursName. Although current audio–visual LLMs perform well in single-modality tasks (e.g., speech recognition), they cannot reliably integrate useful information across modalities. Our findings reveal a fundamental limitation in audio–visual LLMs for now and highlight the need for more reliable speech–video understanding.
\section{\oursName: Cross-modal Speech-Vision Hallucination Evaluation Benchmark}
\label{sec:method}

In this section, we first introduce the taxonomy of our \oursName{} benchmark in Section~\ref{subsec:taxonomy}. We then present our investigation of speech-vision hallucinations from two critical and complementary aspects:  semantic hallucination in Section~\ref{subsec:semantic_halluc} and temporal hallucination in Section~\ref{subsec:temporal_halluc}. Finally, we introduce the dataset creation pipeline in Section~\ref{subsec:dataset_curation} and dataset statistics in Section~\ref{subsec:dataset_statistics}.

\subsection{Task Taxonomy}
\label{subsec:taxonomy}

\noindent\textbf{Overview.} Our work proposes \oursName, the first benchmark that systematically evaluates speech–vision hallucination in audio-visual LLMs.  \oursName{} examines model abilities and reveals hallucination modes from two critical and complementary aspects as follows.  (1) \textbf{Semantic hallucination}:   \textit{what} content in the speech semantically correspond to visual evidence; (2) \textbf{Temporal hallucination}: \textit{when} the narrated events visually occur compared to moment when speech is heard. These two dimensions capture fundamental properties of human speech, including rich semantic information and temporal structures. For each dimension, we design three complementary diagnostic tasks that probe the model in a coarse-to-fine manner, uncovering hallucination modes unexplored in previous benchmarks~\cite{leng2025the,sung-bin2025avhbench,radevski2025dave}.

\noindent\textbf{Task formulation.} Following prior benchmarks~\cite{leng2025the,sung-bin2025avhbench,radevski2025dave}, we adopt a unified question-based formulation for all tasks.  Given an input video paired with speech, we construct text-format questions and ask models to generate corresponding answers based on the video input. We design multiple questions to probe different model behaviors. For consistent evaluation and statistical analysis, we follow prior hallucination benchmarks~\cite{leng2025the,sung-bin2025avhbench,radevski2025dave,yebin2024beaf} to present the questions in either two-choice or multiple-choice formats.  We introduce how these questions are constructed in Section~\ref{subsec:semantic_halluc} and Section~\ref{subsec:temporal_halluc}. We perform rigorous evaluation using multiple metrics, with details in Section~\ref{subsec:exp_setups}.

\subsection{Speech-Vision Semantic Hallucination}
\label{subsec:semantic_halluc}

\noindent\textbf{Motivation.} Existing audio-visual hallucination benchmarks~\cite{leng2025the,sung-bin2025avhbench,radevski2025dave} primarily use environmental sounds (e.g., dog barking and sirens) to indicate the occurrence of a single event.  In contrast, human speech carries rich and fundamental different semantics (e.g., objects or actions), yet remains unexplored.  Importantly, speech does not always describe the actual visual scene, including diverse information such as events that are not visible or even totally unrelated to visual content.  While humans can easily find the correspondence between speech content and relevant visual evidence, it remains unexplored whether current audio-visual LLMs have this ability. To address this, we design a set of diagnostic tasks as follows.

\noindent\textbf{Task definition.} Given a video paired with human speech, we evaluate whether audio-visual LLMs can correctly identify the correspondence between speech content and visual scene. Specifically, a successful model should detect semantic inconsistencies between speech and visual signals, while avoiding hallucinations induced by mismatched or irrelevant speech.

\noindent\textbf{Task design.} We design three complementary coarse-to-fine tasks as follows:  
\begin{enumerate}
    \item \textit{\textbf{Global semantic alignment (GSA):}} 
This question evaluates whether the model can identify if the entire speech semantically matches the visual scene:
\begin{quote}
\small
\textit{Does the speech describe the visual events in the video? (A) Yes (B) No}
\end{quote}
A model succeeds in this task if it chooses ``yes'' for matched samples while ``no'' for mismatched ones. This question examines whether the model can detect the semantic misalignment between speech and vision, rather than assuming alignment. 

\item \textit{\textbf{Fine-grained semantic alignment (FGSA):}}
This question examines whether specific objects mentioned in the speech are visually present in the video:
\begin{quote}
\small
\textit{Is the [object] visible in the video? (A) Yes (B) No
}
\end{quote}
For each sample, the correct answer is ``yes'' if the narrated object visually appears in the video, otherwise ``no''. This task aims to evaluate whether the model can identify fine-grained misalignment among speech and video or hallucinate non-existent entities without grounding them in visual evidence. 

\item \textit{\textbf{Cross-modal semantic binding (CMSB):}}
This question tests whether the model hallucinates an event by incorrectly combining actions mentioned in the speech with objects visible in the video (or vice versa).
\begin{quote}
\small
\textit{Is the following event visible in the video: [event]? (A) Yes (B) No}
\end{quote}
For these composed events, the correct answer is ``no'' since they never occur visually in the video. In this case, a ``yes'' response indicates that the model mistakenly binds speech and visual content to hallucinate non-existent events.

\end{enumerate}

Overall, these three tasks provide a comprehensive evaluation of the model's ability to identify semantic misalignment between speech and visual scenes.

\subsection{Speech-Vision Temporal Hallucination}
\label{subsec:temporal_halluc}

\noindent\textbf{Motivation.} Existing hallucination benchmark~\cite{radevski2025dave}  uses environmental sounds to indicate the current moment, asking questions like ``What is the person doing when siren is heard?''.  This design mainly examines the temporal grounding ability of audio-visual LLMs.  In contrast, speech conveys rich temporal structures. It may describe actions or events not happening in the current moment, but in the past or future.  These properties introduce complex temporal relationships between speech content and visual scenes, which is unexplored in prior studies.

\noindent\textbf{Task definition.} Given a video paired with human speech, we evaluate whether audio–visual LLMs can correctly identify \textit{when} the narrated events visually occur compared to the moment when the speech is heard.  Specifically, a successful model should distinguish whether the speech describes events that happen in the \textit{past}, \textit{present}, or \textit{future}, instead of hallucinating events occurring in an incorrect time.

\noindent\textbf{Task Design.} We design three complementary coarse-to-fine tasks as follows:  

\begin{enumerate}
    \item \textit{\textbf{Temporal alignment (TA).}}  This task aims to evaluate whether the models can identify whether the narrated events are visually happening at the same time when speech is heard:
\begin{quote}
\small
\textit{When the speech is heard, is the described event happening at the same time in the video? (A) Yes (B) No}
\end{quote}
To correctly answer this question, the model needs to first understand the speech to capture narrated events, then identify the events visually happening in the video, and finally evaluate whether they align in time.  

\item \textit{\textbf{Temporal forecasting (TF).}} 
This question evaluates whether the model can identify the relative temporal position of narrated events along the video timeline: 
\begin{quote}
\small
\textit{Relative to the moment when the speech is heard, when does the narrated event occur in the video? (A) Past (B) Present (C) Future
}
\end{quote}
This task evaluates the ability of models not only in cross-modality reasoning, but also in fine-grained reasoning over time.

\item \textit{\textbf{Cross-modal temporal binding (CMTB).}} 
This question evaluates whether the model can still be grounded in visual evidence while the speech describes an event happening at a different time.
\begin{quote}
\small
\textit{When the speech is heard, what visual action is occurring in the video? (A) [Speech-mentioned event that happens in a different time] (B) [Visual event that is happening when the person is speaking] (C) [A distracting event from other videos] }
\end{quote}
To correctly answer this question, the models need to identify the actions actually happening visually and avoid hallucinating the narrated events into the wrong temporal position.
\end{enumerate}

Overall, these three tasks evaluate the temporal understanding abilities of audio-visual LLMs across modalities. 

\subsection{Dataset Construction}
\label{subsec:dataset_curation}
To systematically study speech-vision hallucinations in audio-visual LLMs, we propose an \textit{automated dataset construction pipeline with human verification.} Specifically, our pipeline transforms well-aligned speech-video pairs ($A$, $V$) into controlled variations that correspond to semantic or temporal conditions.  For clarity, for two-choice questions, we term samples with ground-truth labels ``yes'' as positive samples, while those labeled ``no'' as negative samples.

\noindent\textbf{Video collection and preprocessing.} We collect synchronized speech-video pairs from the widely-applied YouCook2 dataset~\cite {zhou2018towards}, where the narrations well align with visual actions. We follow prior work~\cite{Kim_2025_CVPR,Zhou_2024_CVPR,schiappa2022robustness} to adopt the validation set from YouCook2 for evaluation. Based on the annotations from YouCook2~\cite {zhou2018towards}, we clip each video into procedure segments. We then use an automatic speech recognition (ASR) model, the widely-used Whisper model~\cite{pmlr-v202-radford23a}, to obtain the speech transcript for each video.

\noindent\textbf{Data construction for semantic hallucination tasks.} We introduce how to obtain samples for each task as follows.
\begin{enumerate}
    \item \textit{Global semantic alignment (GSA):} Positive samples for GSA task are those where speech content correctly describes visual scenes. Therefore, we use the original well-aligned speech-video pair ($A_i$, $V_i$) from YouCook2~\cite {zhou2018towards} dataset as positive samples, where $i$ indicates the index for the original video clip. To create negative samples that require speech-video misalignment, we randomly sample another video $j$ and pair its audio $A_j$ with the current video $V_i$, forming a disturbed pair ($A_j$, $V_i$)  that breaks speech-vision correspondence. Note that both the aligned speech-video pairs ($A_i$, $V_i$) and the disturbed ones ($A_j$, $V_i$) all serve as video candidates for other semantic hallucination tasks.
    \item \textit{Fine-grained semantic alignment (FGSA):} We prompt GPT model~\cite{gpt5} to find the visible and invisible objects in the video based on detailed video and audio annotations from YouCook2~\cite {zhou2018towards} dataset. We then fill the question template with visible objects to create positive samples, and with narrated but invisible objects as negative samples.
    \item \textit{Cross-modal semantic binding (CMSB):} We use GPT to extract the visible action–object pairs based on video annotations as positive event pairs. To create negative samples, we first use GPT to extract speech-only actions, i.e., actions that are mentioned in speech but invisible in videos. Similarly, we extract speech-only objects, vision-only actions, and vision-only objects. We then create cross-modal action-object combinations as follows: (speech-only action, vision-only object) and (vision-only action, speech-only object). Note that these combinations do not occur visually in the video. To ensure data quality, we further use GPT to filter out unrealistic combinations and keep only reasonable events as negative samples.
\end{enumerate}

\noindent \textbf{Data construction for temporal hallucination tasks.} We construct video samples based on annotations about the start and end time of events from YouCook2~\cite {zhou2018towards} dataset. For each video, we identify the time around the moment when speech is heard as $t_{speak}$. We then define the time when narrated events visually occur as $t_{visual}$. Given a speech-video pair, if $t_{speak}$ is close to  $t_{visual}$, we define it as temporally aligned, i.e., when the person is speaking, the narrated events are visually happening in the \textit{present} moment. In the case that $t_{speak}$ and $t_{visual}$ are far from each other, we define them as temporally-misaligned. Specifically, if $t_{visual}$ is much smaller than $t_{speak}$, we define that the narrated events happen in the \textit{past} moment, and \textit{future} moment if  $t_{visual}$ is much larger than $t_{speak}$. For cross-modal temporal binding task, we adopt event annotations from YouCook2~\cite {zhou2018towards} as option candidates.

\noindent \textbf{Quality control.} To ensure dataset quality, we first use GPT model to filter out unqualified samples, such as videos where speech is too short or does not clearly describe visual scenes. We then perform human-in-the-loop for final verification, ensuring all the samples are clear. 

\subsection{Dataset Statistics}
\label{subsec:dataset_statistics}

\begin{figure}[t!]
    \centering
    \includegraphics[width=0.95\linewidth]{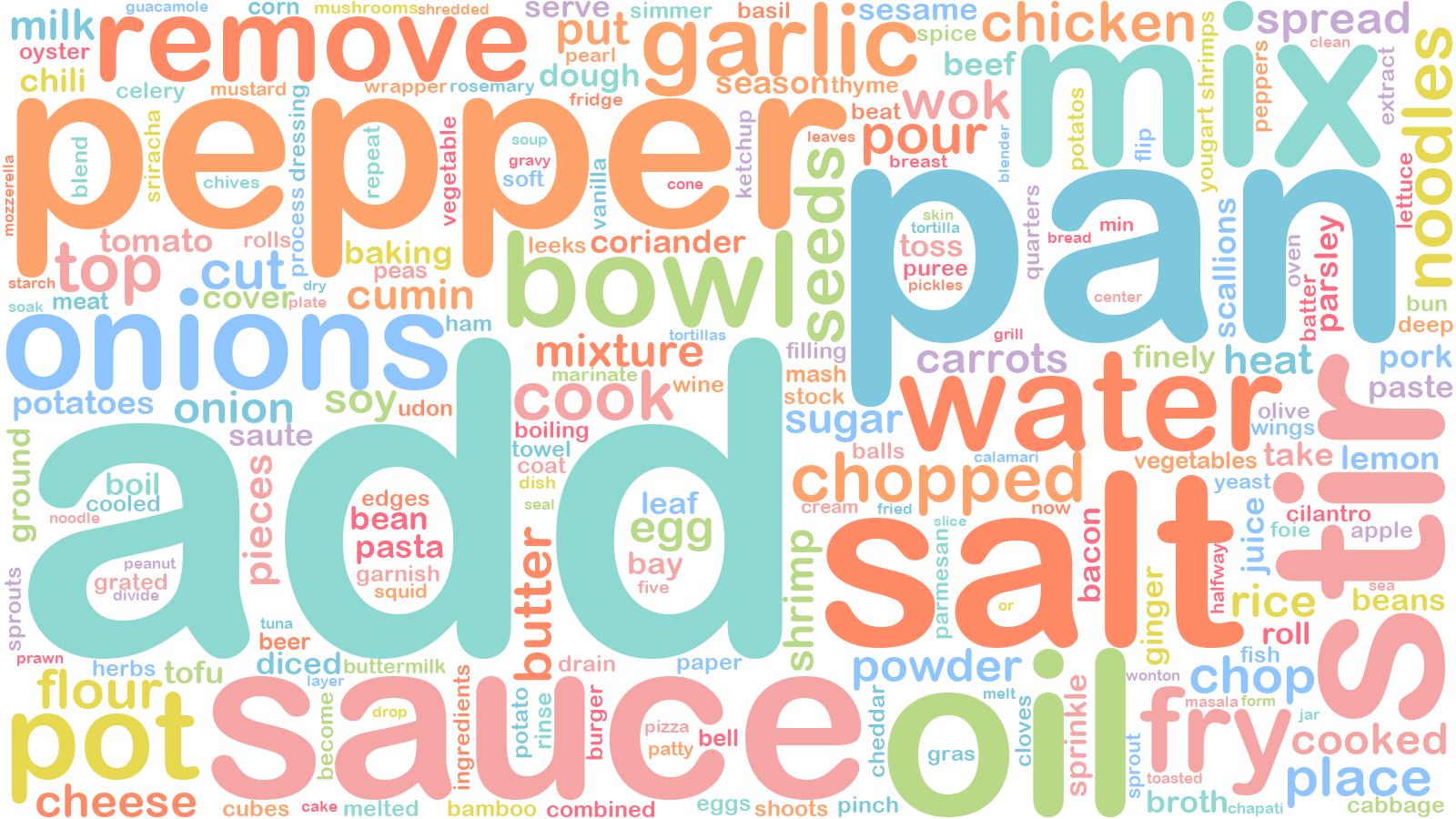}
    \vspace{-2mm}
    \vspace{-1mm}
    \caption{
        \textbf{Distributions of actions and objects from \oursName.} 
    }
    \label{fig:wordclouds}
\end{figure}

Our \oursName{} benchmark contains 2405 video-question pairs, including 1422 samples for semantic hallucination tasks and 983 pairs for temporal hallucination. We show the sample distribution for each task in Table~\ref{tab:unified_task_distribution}. For each task, we ensure a balanced number of samples for each option to ensure fair evaluation. We also show the distribution of action and objects in Figure~\ref{fig:wordclouds}.

\begin{table}[t]
\centering
\small
\resizebox{1\linewidth}{!}{
\begin{tabular}{lccc}
\toprule
\textbf{Category} & \textbf{Task} & \textbf{Abbrev.} & \textbf{Count} \\
\midrule
\multirow{3}{*}{Semantic}
 & Global Semantic Alignment & GSA  & 444 \\
& Fine-Grained Semantic Alignment & FGSA  &444 \\
 & Cross-Modal Semantic Binding & CMSB  & 534 \\
\cmidrule(lr){2-4}
 & \textbf{Semantic Hallucination Total} &  & \textbf{1422} \\
\midrule
\multirow{3}{*}{Temporal}
 & Temporal Alignment   & TA  &  368 \\
 & Temporal Forecasting & TF  &  276 \\
 & Cross-Modal Temporal Binding& CMTB & 339 \\
\cmidrule(lr){2-4}
 & \textbf{Temporal Hallucination Total} &  & \textbf{983} \\
\midrule
\textbf{Overall Total} & & & \textbf{2405} \\
\bottomrule
\end{tabular}}
\caption{\textbf{Task taxonomy.} We summarize the tasks in \oursName{} as well as their abbreviations. Our benchmark includes 2405 video-question pairs on 6 different speech-vision hallucination tasks.}
\label{tab:unified_task_distribution}
\end{table}

\section{Experiments}
\label{sec:experiments}

In this section, we first introduce our experiment setups in Section~\ref{subsec:exp_setups}, then report the results of state-of-the-art audio-visual LLMs in Section~\ref{subsec:results}. We further analyze the underlying reasons for model failures and provide insights for model improvement in Section~\ref{subsec:analysis}.

\subsection{Experiment Setup}
\label{subsec:exp_setups}

\noindent\textbf{Models.} We evaluate our \oursName{} benchmark on six state-of-the-art audio-visual LLMs, including both open-source and commercial ones. These models have achieved remarkable success and are widely used for omni-modal understanding tasks. For open-source models, we evaluate the latest Qwen3-Omni~\cite{xu2025qwen3}, Qwen2.5-Omni~\cite{xu2025qwen25omnitechnicalreport}, video-SALMONN 2~\cite{tang2025video}, and VideoLLaMA 2~\cite{cheng2024videollama}. For commercial models, we evaluate the latest Gemini 2.5 Pro. For different tasks, we prompt the model with different questions as discussed in Section~\ref{sec:method}. We evaluate all models in a zero-shot manner, without additional model training.

\noindent\textbf{Evaluation metrics.} We follow previous audio-visual benchmarks~\cite{sung-bin2025avhbench,Chowdhury_2025_ICCV,radevski2025dave} to design our tasks in a two-choice or multiple-choice manner. For evaluation, we use standard metrics widely used in existing studies~\cite{sung-bin2025avhbench,Chowdhury_2025_ICCV,radevski2025dave}. For two-choice tasks, we report accuracy, precision, recall, and F1 score. Specifically, accuracy is defined as the ratio of correctly predicted samples over all samples. Precision is the ratio of true positives to all predicted positives, while recall is the ratio of true positives to all true positives. To evaluate the case when precision and recall diverge, we use F1 score, which is defined as the harmonic mean of precision and recall. We take samples with the "yes" option as positive ones when computing the precision, recall, and F1 metrics.  We report accuracy over all samples for multiple-choice tasks.

\subsection{Results}
\label{subsec:results}

We evaluate \oursName{} on six state-of-the-art audio-visual LLMs, including the recent open-source model Qwen3-Omni~\cite{xu2025qwen3} and Gemini 2.5 Pro. We report the results for semantic hallucination and temporal hallucination tasks in Tables~\ref{tab:semantic_results} and \ref{tab:temporal_results}, respectively. Note that our dataset is option-balanced, and we report the result of random choices in each task for reference.

\noindent\textbf{Semantic hallucination.} Semantic hallucination tasks evaluate whether the model can identify what speech content is supported by visual evidence.  As shown in Table~\ref{tab:semantic_results}, all open-source models show severe failure modes in our benchmark. For the \textit{global semantic alignment (GSA)} task, most models achieve an accuracy of around 50\%, close to random guess results. This indicates that models tend to believe that the speech describes visual scenes, without really reasoning about their alignment. Therefore, even in the cases that speech and video describe two distinctly different scenes, e.g., boiling noodles and frying meat patties, models still choose the positive alignment answer, leading to a high recall but low precision. Our GSA task reveals a high alignment bias in existing models from the global semantics perspective. Notably, Gemini 2.5 Pro achieves a high accuracy of 93.10\%, demonstrating a huge gap, and also highlights that our dataset is solvable. In the \textit{fine-grained semantic alignment (FGSA)} task,  video-SALMONN~\cite{pmlr-v235-sun24l} achieves a poor accuracy, while Qwen2.5-Omni~\cite{xu2025qwen25omnitechnicalreport} and Qwen3-Omni~\cite{yang2025qwen3technicalreport} show a significant improvement, comparable to Gemini 2.5 Pro. These results show that Qwen2.5-Omni~\cite{xu2025qwen25omnitechnicalreport} and Qwen3-Omni~\cite{yang2025qwen3technicalreport} have impressive fine-grained cross-modal grounding ability compared to other models. For the \textit{cross-modal semantic binding (CMSB)} task, all models perform worse than the FGSA task, highlighting the challenge to the models. Note that for the CMSB task, the models not only need to identify whether the object is visible, but also need to recognize the absence of actions. This task also evaluates the model's ability in whether they can bind the action with its corresponding objects and vice versa.

\noindent \textbf{Temporal hallucination.} Temporal hallucination tasks evaluate whether the models can determine if the narrated events visually occur at the \textit{same} time when the speech is heard, or at a \textit{different} time (e.g., past or future). As shown in Table~\ref{tab:temporal_results}, all open-source models perform poorly at temporal alignment (TA) and temporal forecasting (TF) tasks, with an accuracy around the random choices.  These results demonstrate that existing open-source models fail to find the correct temporal correspondence between speech content and visual events, a critical ability for audio-visual LLMs but remains unexplored in previous studies. Interestingly,  Qwen3-Omni~\cite{xu2025qwen3} achieves significant improvement on the cross-modal temporal binding (CMTB) task compared to prior open-source models, where speech acts as a distractor for visual recognition since it describes events at a different time.  In contrast, Gemini-2.5 Pro significantly outperforms all open-source models in all temporal tasks. These results demonstrate that our \oursName{} benchmark effectively uncovers a new and important set of temporal understanding failures in the open-source audio-visual LLMs, and suggests a direction for future improvement. 

\subsection{Analysis}
\label{subsec:analysis}

\begin{table*}[t]
\centering
\caption{\textbf{Results for semantic hallucination tasks.}
We evaluate state-of-the-art audio-visual LLMs on \oursName{}, including both open-source model (e.g.,Qwen3-Omni~\cite{xu2025qwen3}) and commercial models (e.g.,Gemini 2.5 Pro~\cite{Gemini2.5pro}). Results across three semantic hallucination tasks show that open-source models perform poorly in \oursName{}, e.g., most models achieve an accuracy near random choice on GSA task. In contrast, Gemini 2.5 Pro achieves a strong performance, highlighting the gap between open-source and commercial models in speech-vision understanding.
}
\label{tab:semantic_results}

\resizebox{\textwidth}{!}{
\begin{tabular}{lcccccccccccc}
\toprule

\multicolumn{13}{c}{\textbf{Semantic Hallucination}} \\
\midrule

\textbf{Model} &
\multicolumn{4}{c}{\textbf{Global semantic alignment (GSA)}} &
\multicolumn{4}{c}{\textbf{Fine-grained semantic alignment (FGSA)}} &
\multicolumn{4}{c}{\textbf{Cross-modal semantic binding (CMSB)}} \\
\cmidrule(lr){2-5}\cmidrule(lr){6-9}\cmidrule(lr){10-13}
& Acc. ($\uparrow$) & Prec. ($\uparrow$) & Recall ($\uparrow$) & F1($\uparrow$) &
  Acc. ($\uparrow$) & Prec. ($\uparrow$) & Recall ($\uparrow$) & F1($\uparrow$) &
  Acc. ($\uparrow$) & Prec. ($\uparrow$) & Recall ($\uparrow$) & F1($\uparrow$) \\
\midrule

Gemini 2.5 Pro~\cite{Gemini2.5pro} &
93.10 & 90.24 & 96.64 & 93.33 &
86.06 & 80.22 & 95.79 & 87.32 &
78.56 & 74.58 & 92.98 & 82.77 \\
\midrule
video-SALMONN~\cite{pmlr-v235-sun24l} &
52.05 & 51.06 & 99.25 & 67.43 &
49.84 & 49.92 & 96.12 & 65.71&
61.10 & 56.63& 94.88& 70.92 \\
video-SALMONN 2~\cite{tang2025video} &
53.92 & 62.35 & 19.78 & 30.03 &
79.58 & 79.42 & 79.94 & 79.68 &
72.77 & 65.71 & 95.26 & 77.77\\

VideoLLaMA 2~\cite{cheng2024videollama} &
50.00&50.00&100.00&66.67&67.57&58.44&94.74&72.29&73.03&68.22&96.69&80.00 \\

Qwen2.5-Omni~\cite{xu2025qwen25omnitechnicalreport} &77.70&73.75&86.04&79.42&81.08&77.60&87.39&82.20&72.28&64.48&99.25&78.17 \\

Qwen3-Omni~\cite{xu2025qwen3} &55.10&52.86&100.00&69.16&79.50&72.00&97.30&82.76&74.34&66.17&99.63&79.52 \\

\midrule

Random Choice & 50.00 & 50.00 & 50.00 & 50.00 &
50.00 & 50.00 & 50.00 & 50.00 &
50.00 & 50.00 & 50.00 & 50.00 \\

\bottomrule
\end{tabular}
}
\end{table*}

\begin{table}[t]
\centering
\caption{\textbf{Results for temporal hallucination tasks.} We report the results of state-of-the-art models on temporal hallucination tasks in \oursName{}. All open-source models achieve a near-random performance, while Gemini 2.5 Pro~\cite{Gemini2.5pro} significantly outperforms others. These results highlight the need for open-source model improvement in speech-vision understanding.
}
\label{tab:temporal_results}

\resizebox{\linewidth}{!}{
\begin{tabular}{lcccccccc}
\toprule

\multicolumn{7}{c}{\textbf{Temporal Hallucination}} \\
\midrule

\textbf{Model} &
\multicolumn{4}{c}{\textbf{TA}} &
\multicolumn{1}{c}{\textbf{TF}} &
\multicolumn{1}{c}{\textbf{CMTB}} \\
\cmidrule(lr){2-5} \cmidrule(lr){6-6} \cmidrule(lr){7-7}
& Acc. ($\uparrow$) & Prec. ($\uparrow$) & Recall ($\uparrow$) & F1($\uparrow$) &
  Acc. ($\uparrow$)  &
  Acc. ($\uparrow$)  \\
\midrule

Gemini 2.5 Pro~\cite{Gemini2.5pro} & 
 85.17&82.67&89.00&85.71&53.89&69.25
\\

\midrule

video-SALMONN~\cite{pmlr-v235-sun24l} 
& 50.00 & 50.00 &87.08 & 63.53& 32.81 & 37.39
\\

video-SALMONN 2~\cite{tang2025video}
& 50.00 & 50.00 & 100.00 & 66.67 & 33.33 & 48.58
\\

VideoLLaMA 2~\cite{cheng2024videollama} &50.00&50.00&100.00&66.67&32.25&45.72 \\

Qwen2.5-Omni~\cite{xu2025qwen25omnitechnicalreport} &50.27&50.28&49.46&49.86&31.52&53.10 \\

Qwen3-Omni~\cite{xu2025qwen3} &51.11&50.84&100.00&67.41&30.60&61.75\\

\midrule
Random Choice & 50.00 & 50.00 & 50.00 & 50.00 &
33.33&
33.33 \\

\bottomrule
\end{tabular}
}
\end{table}

\begin{table*}[t]
\centering
{
\caption{\textbf{Results for analysis experiments.} To analyze the reason why open-source models fail in \oursName{}, we perform experiments of Qwen-3-Omni~\cite{yang2025qwen3technicalreport} in different experiment conditions and report the accuracy. Our results show that adding speech transcripts can improve accuracy in several tasks (e.g., FGSA). We also present how different modality inputs influence the accuracy in different tasks.
}
\label{tab:analysis_results}
\resizebox{0.9\textwidth}{!}{
\begin{tabular}{lccccccccc}
\toprule
\textbf{Model} &  
\multicolumn{4}{c}{\textbf{Semantic Hallucination}} 
& \multicolumn{4}{c}{\textbf{Temporal Hallucination}} 
\\
\cmidrule(lr){2-5} \cmidrule(lr){6-9} 
  & GSA ($\uparrow$) &  FGSA ($\uparrow$) & CMSB ($\uparrow$) & Average ($\uparrow$)
 & TA ($\uparrow$) & TF ($\uparrow$) & CMTB ($\uparrow$) & Average ($\uparrow$)
 \\
\midrule

Original Qwen3-Omni~\cite{yang2025qwen3technicalreport}  &55.10&79.50&74.34&69.64& 51.11&30.60&61.75 & 47.82 \\
\midrule 
With speech transcript &53.29&83.14&76.40&70.94 &50.56&33.21&49.70 &  44.49\\
\midrule
Video-only input &49.77&84.91&87.08 & 73.92&50.27&33.21&47.18 &43.55\\
Audio-only input &50.45&49.32&80.34&60.03&50.42&24.54&50.45&41.80\\

\midrule
Random Choice    & 50.00 & 50.00 & 50.00 & 50.00 & 50.00 & 33.33 & 33.33 & 38.88  \\
\bottomrule
\end{tabular}
}
}
\end{table*}

Section~\ref{subsec:results} demonstrates that our work uncovers new and critical failure modes of advanced audio-visual LLMs when understanding video-speech pairs. In this section, we provide a comprehensive analysis of why current models hallucinate on \oursName{}. Our analysis provides an in-depth understanding of how each component in audio-visual LLMs works, and offers insights for future model improvement.

\noindent\textbf{Do models recognize what people are saying in the audio?}
To analyze whether speech recognition ability is the reason for the model failures in  \oursName{}, we evaluate Qwen3-Omni~\cite{xu2025qwen3}, the state-of-the-art model, on the automatic speech recognition (ASR) task. Specifically, given each video in \oursName{}, we prompt the models with the following question: \textit{What does the person say in the audio?}. We record the model outputs as predicted speech content of Qwen3-Omni. For quantitative evaluation, we use a widely-applied Whisper model~\cite{pmlr-v202-radford23a} to generate the pseudo ground-truth labels. Experimental results show that Qwen3-Omni achieves a low word error rate (WER) of 0.0915 and a high word information preserved (WIP) of 0.8863.
These results demonstrate that Qwen3-Omni~\cite{xu2025qwen3} could recognize most of the speech contents in \oursName{}, indicating that speech recognition ability is not the key bottleneck for the poor performance of audio-visual LLMs on \oursName{}.

\noindent \textbf{Can language cues mitigate speech-vision hallucinations?} Existing audio-visual LLMs typically encode the visual and audio inputs to the text token space of LLMs, using LLMs to perform multimodal reasoning. However, this design may not align non-text modalities and text well, leading to suboptimal understanding of multimodal information. To evaluate whether poor results in \oursName{} come from the suboptimal representation of audio tokens, we explicitly use the speech transcript as additional text input to the model. Specifically, we use Whisper model~\cite{pmlr-v202-radford23a} to transcribe the speech, and concatenate them with the original text prompt (e.g., questions) as inputs. Table~\ref{tab:analysis_results} shows that adding speech transcripts improves the model's performance on FGSA and CMSB, indicating that speech content in the text format helps the model have a better understanding. For the CMTB task, speech acts as distractors by describing events that happen at a different time. This aligns well with the results that adding speech transcripts leads to more severe hallucination on CMTB, indicated by a performance drop. Notably, using speech transcripts achieves comparable results in GSA, TA, and TF tasks, indicating that those tasks provide more complex challenges to audio-visual LLMs beyond speech recognition.

\noindent \textbf{Can models recognize the time when the speech is heard?} Temporal hallucination tasks evaluate whether models can identify the temporal correspondence between narrated events and visual scenes. To understand the model's ability in recognizing time, we explicitly ask the model to output the start and end times when the speech is heard. Specifically, we prompt the models with the following question: \textit{``When is the speech heard in the audio? Provide the start and end timestamps (in seconds)."} We evaluate Qwen3-Omni~\cite{xu2025qwen3} on our \oursName{} and compare the outputs with predictions of Silero-VAD~\cite{silero_vad2024}, a strong and widely used voice activity detection model. For quantitative evaluation, we use predictions of Silero-VAD as pseudo-labels and compute the mIoU between Silero-VAD and Qwen3-Omni, obtaining a mIoU of 0.8844. This indicates that Qwen3-Omni has a strong temporal localization ability of speech segments even in our challenging video scenarios. Therefore, the ability to identify when speech is heard is not the main factor that leads to the poor performance on our \oursName{}. Our analysis also demonstrates that \oursName{} evaluates a broader ability beyond speech temporal localization.

\noindent \textbf{Do models correctly use both video and speech modality for comprehension?} Our \oursName{} evaluates the cross-modality understanding ability of existing audio-visual LLMs. To examine whether models can correctly use video and speech and how much they rely on each modality in different tasks, we evaluate Qwen3-Omni~\cite{xu2025qwen3} by removing one modality at the input level. Table~\ref{tab:analysis_results} shows that when taking speech-only inputs, accuracy on FGSA, CMSB, and CMTB drops significantly, indicating effective use of the visual content. In contrast, using video-only inputs even improves FGSA and CMSB since speech acts as a distractor in these tasks. The performance on CMTB drops since it requires speech for temporal grounding. Removing either modality has a limited impact on GSA, TA, and TF tasks, revealing a limitation of current audio-visual LLMs in identifying the alignment between speech and video.

\noindent \textbf{Summary.} In this section, we analyze the reason for the poor performance of current audio-visual LLMs in our \oursName{}. Our study shows that current models perform well in single-modality tasks, such as speech recognition. However, they struggle to fully integrate information from both video and speech for multimodal reasoning.  Our findings highlight that \oursName{} provides significant challenges for current models and the urgent need for speech-grounded video understanding.
\section{Related Work}
\label{sec:related_work}

\noindent \textbf{Multimodal Large Language Models (MLLMs).} Recently, 
LLMs have been extended to support inputs from multiple modalities~\cite{Chen_2024_CVPR,li2025llavaonevision,lin-etal-2024-video}, and are widely used in multiple applications, such as multimodal embedding ~\cite{Kim_2026_WACV}, image retouching~\cite{yebin2025retouchllmtrainingfreecodebasedimage}, and automatic model discovery~\cite{jung-mok2025automated}. As a milestone work for language-image modeling, LLaVa~\cite{liu2023visual} connects CLIP~\cite{pmlr-v139-radford21a} vision encoder with language decoder Vicuna~\cite{vicuna2023}. Specifically,   LLaVa~\cite{liu2023visual}  projects vision representations to the embedding space of language tokens. The language encoder then performs multimodal reasoning based on both vision and language tokens. This framework has been widely-applied in the advancement of MLLMs~\cite{li2025llavaonevision, li2023llavamed,Liu_2024_CVPR}, such as Video-LLaVA~\cite{lin-etal-2024-video} for video understanding. On the other hand, audios convey rich information and plays a critical role for humans when understanding the world.
Recent MLLMs have extended to include audio signals for omnimodal perception and reasoning~\cite{cheng2024videollama, tang2025video,xu2025qwen25omnitechnicalreport,xu2025qwen3}, such as OneLLM~\cite{han2024onellm}, Video-LLaMa~\cite{zhang-etal-2023-video}, and Video-LLaMA2~\cite{cheng2024videollama}. Specifically, Qwen2.5-Omni~\cite{xu2025qwen25omnitechnicalreport} and Qwen3-Omni~\cite{xu2025qwen3} achieve great attention due to their impressive performance on downstream tasks~\cite{sakshi2025mmau,hong2026worldsense}. In this work, we focus on audio-visual LLMs, which incorporate text,  video, and audio inputs for a unified understanding. 

\noindent \textbf{Evaluating Multimodal Large Language Models.} Prior studies have specifically designed multiple benchmarks to evaluate the ability of multimodal LLMs~\cite{Chowdhury_2025_ICCV,gong2025avodyssey,hong2026worldsense,cheng2025videoholmesmllmthinklike}.  LENS~\cite{hyeon-woo2025vlms} is designed to first instruct vision-language models then check its readiness.  AV-Odyssey Bench~\cite{gong2025avodyssey} finds that audio-visual LLMs often fail in simple tasks that humans can easily identify, such as determining which of two sounds is louder. AVTRUSTBENCH~\cite{Chowdhury_2025_ICCV} focuses on model robustness, including tasks like adversarial attack and compositional reasoning. Specifically, it evaluates whether the model could identify the case where there are no correct options. WorldSense~\cite{hong2026worldsense} examines the omnimodal understanding ability by designing questions based on both video and audio. Video-Holmes~\cite{cheng2025videoholmesmllmthinklike} includes videos from suspense short films to test the reasoning ability of models. Another work SMILE~\cite{hyun-etal-2024-smile} investigates the reason why people laugh in videos, while proposing a strong baseline leveraging the reasoning ability of text-only LLMs. In contrast, our work systematically investigates the hallucination problem of audio-visual LLMs, specifically the speech-video hallucination. 

\noindent \textbf{Hallucinations in MLLMs.} Hallucination is a key challenge to MLLMs, where the models produce plausible responses but are not grounded in the input signals~\cite{zhu2025alleviating,zou2025look}. Early studies focus on vision-language hallucination~\cite{li-etal-2023-evaluating,hu2023ciemcontrastiveinstructionevaluation,zou2025look}, introducing benchmarks to evaluate whether models can correctly identify the visual presence of objects, such as BEAF~\cite{yebin2024beaf}, POPE~\cite{li-etal-2023-evaluating}, and CIEM~\cite{hu2023ciemcontrastiveinstructionevaluation}. Recently, a branch of studies~\cite{sung-bin2025avhbench,leng2025the} reveals that audio-visual LLMs also hallucinate given the multimodal video-audio inputs. For example, AVHBench~\cite{sung-bin2025avhbench} finds that audio-visual LLMs often assume a visible object is making sound without referring to the audio evidence, while another work~\cite{leng2025the} investigates how spurious inter-modality correlations influence the hallucination behaviors. Despite these investigations, existing audio-visual hallucination benchmarks typically investigate the environmental sound (e.g., dog barking or sirens) and use them to indicate the occurrence of events. In contrast, speech conveys fundamentally different and rich information, yet it is unexplored audio-visual LLMs could correctly align speech content with visual evidence. Our work fills this gap and provides a systematic investigation of speech-vision hallucination in audio-visual LLMs.
\section{Conclusion}
\label{sec:conclusion}

In this work, we study speech-vision hallucinations in audio-visual LLMs, which is unexplored in prior work. To systematically address this, we propose \oursName{}, a comprehensive benchmark that investigates speech-vision hallucination from two critical and complementary aspects: semantic and temporal. Specifically, we design three coarse-to-fine tasks for each aspect, resulting in six tasks in total. Experimental results show that state-of-the-art open-source models perform poorly on our \oursName{}, achieving near-random accuracy on multiple tasks.  Gemini-2.5 Pro significantly outperforms the open-source models, highlighting a huge gap between open-source and commercial models. We further analyze the reason for the failure of open-source models. Our analysis suggests that current models show good ability in single-modality tasks, such as speech recognition and temporal localization. However, these models fail in leveraging and integrating information across multiple modalities. Our work uncovers a critical limitation of existing audio-visual LLMs and highlight the need for reliable speech-video understanding.

\section*{Acknowledgment}

This work was supported by the  InnoCORE program of the Ministry of Science and ICT (N10250156, KAIST InnoCore LLM) (23.3\%); the National Research Foundation of Korea(NRF) funded by the Korea government(MSIT) (No.~RS-2024-00451947) (30\%); and the Institute of Information \& communications Technology Planning \& Evaluation (IITP) grant funded by the Korea government(MSIT) (No.~RS-2024-00457882 (23.3\%), National AI Research Lab Project; No. 2022-0-00124; No.~2022-0-00124, No.~RS-2022-II220124 (23.3\%), Development of Artificial Intelligence Technology for Self-Improving Competency-Aware Learning). 

{
    \small
    \bibliographystyle{ieeenat_fullname}
    \bibliography{main}

\begin{thebibliography}{46}
\providecommand{\natexlab}[1]{#1}
\providecommand{\url}[1]{\texttt{#1}}
\expandafter\ifx\csname urlstyle\endcsname\relax
  \providecommand{\doi}[1]{doi: #1}\else
  \providecommand{\doi}{doi: \begingroup \urlstyle{rm}\Url}\fi

\bibitem[Alayrac et~al.(2022)Alayrac, Donahue, Luc, Miech, Barr, Hasson, Lenc, Mensch, Millican, Reynolds, Ring, Rutherford, Cabi, Han, Gong, Samangooei, Monteiro, Menick, Borgeaud, Brock, Nematzadeh, Sharifzadeh, Binkowski, Barreira, Vinyals, Zisserman, and Simonyan]{alayrac2022flamingo}
Jean-Baptiste Alayrac, Jeff Donahue, Pauline Luc, Antoine Miech, Iain Barr, Yana Hasson, Karel Lenc, Arthur Mensch, Katherine Millican, Malcolm Reynolds, Roman Ring, Eliza Rutherford, Serkan Cabi, Tengda Han, Zhitao Gong, Sina Samangooei, Marianne Monteiro, Jacob Menick, Sebastian Borgeaud, Andrew Brock, Aida Nematzadeh, Sahand Sharifzadeh, Mikolaj Binkowski, Ricardo Barreira, Oriol Vinyals, Andrew Zisserman, and Karen Simonyan.
\newblock Flamingo: a visual language model for few-shot learning.
\newblock In \emph{NeurIPS}, 2022.

\bibitem[Chen et~al.(2024)Chen, Wu, Wang, Su, Chen, Xing, Zhong, Zhang, Zhu, Lu, Li, Luo, Lu, Qiao, and Dai]{Chen_2024_CVPR}
Zhe Chen, Jiannan Wu, Wenhai Wang, Weijie Su, Guo Chen, Sen Xing, Muyan Zhong, Qinglong Zhang, Xizhou Zhu, Lewei Lu, Bin Li, Ping Luo, Tong Lu, Yu Qiao, and Jifeng Dai.
\newblock Internvl: Scaling up vision foundation models and aligning for generic visual-linguistic tasks.
\newblock In \emph{CVPR}, pages 24185--24198, 2024.

\bibitem[Cheng et~al.(2025)Cheng, Ge, Wang, Ge, Liao, and Shan]{cheng2025videoholmesmllmthinklike}
Junhao Cheng, Yuying Ge, Teng Wang, Yixiao Ge, Jing Liao, and Ying Shan.
\newblock Video-holmes: Can mllm think like holmes for complex video reasoning?
\newblock \emph{arXiv preprint arXiv:2505.21374}, 2025.

\bibitem[Cheng et~al.(2024)Cheng, Leng, Zhang, Xin, Li, Chen, Zhu, Zhang, Luo, Zhao, and Bing]{cheng2024videollama}
Zesen Cheng, Sicong Leng, Hang Zhang, Yifei Xin, Xin Li, Guanzheng Chen, Yongxin Zhu, Wenqi Zhang, Ziyang Luo, Deli Zhao, and Lidong Bing.
\newblock Videollama 2: Advancing spatial-temporal modeling and audio understanding in video-llms.
\newblock \emph{arXiv preprint arXiv:2406.07476}, 2024.

\bibitem[Chiang et~al.(2023)Chiang, Li, Lin, Sheng, Wu, Zhang, Zheng, Zhuang, Zhuang, Gonzalez, Stoica, and Xing]{vicuna2023}
Wei-Lin Chiang, Zhuohan Li, Zi Lin, Ying Sheng, Zhanghao Wu, Hao Zhang, Lianmin Zheng, Siyuan Zhuang, Yonghao Zhuang, Joseph~E. Gonzalez, Ion Stoica, and Eric~P. Xing.
\newblock Vicuna: An open-source chatbot impressing gpt-4 with 90\%* chatgpt quality, 2023.

\bibitem[Chowdhury et~al.(2025)Chowdhury, Nag, Dasgupta, Wang, Elhoseiny, Gao, and Manocha]{Chowdhury_2025_ICCV}
Sanjoy Chowdhury, Sayan Nag, Subhrajyoti Dasgupta, Yaoting Wang, Mohamed Elhoseiny, Ruohan Gao, and Dinesh Manocha.
\newblock Avtrustbench: Assessing and enhancing reliability and robustness in audio-visual llms.
\newblock In \emph{ICCV}, pages 1590--1601, 2025.

\bibitem[Gong et~al.(2024)Gong, Feng, Li, Wang, Cheng, Yang, Han, Wang, Bai, Yang, and Yue]{gong2025avodyssey}
Kaixiong Gong, Kaituo Feng, Bohao Li, Yibing Wang, Mofan Cheng, Shijia Yang, Jiaming Han, Benyou Wang, Yutong Bai, Zhuoran Yang, and Xiangyu Yue.
\newblock Av-odyssey bench: Can your multimodal llms really understand audio-visual information?
\newblock \emph{arXiv preprint arXiv:2412.02611}, 2024.

\bibitem[Google(2025)]{Gemini2.5pro}
Google.
\newblock Gemini 2.5 pro.
\newblock \url{https://docs.cloud.google.com/vertex-ai/generative-ai/docs/models/gemini/2-5-pro}, 2025.

\bibitem[Han et~al.(2024)Han, Gong, Zhang, Wang, Zhang, Lin, Qiao, Gao, and Yue]{han2024onellm}
Jiaming Han, Kaixiong Gong, Yiyuan Zhang, Jiaqi Wang, Kaipeng Zhang, Dahua Lin, Yu Qiao, Peng Gao, and Xiangyu Yue.
\newblock Onellm: One framework to align all modalities with language.
\newblock In \emph{CVPR}, pages 26584--26595, 2024.

\bibitem[Hong et~al.(2026)Hong, Yan, Cai, Jiang, Hu, and Xie]{hong2026worldsense}
Jack Hong, Shilin Yan, Jiayin Cai, Xiaolong Jiang, Yao Hu, and Weidi Xie.
\newblock Worldsense: Evaluating real-world omnimodal understanding for multimodal {LLM}s.
\newblock In \emph{ICLR}, 2026.

\bibitem[Hu et~al.(2023)Hu, Zhang, Zhao, and Sun]{hu2023ciemcontrastiveinstructionevaluation}
Hongyu Hu, Jiyuan Zhang, Minyi Zhao, and Zhenbang Sun.
\newblock Ciem: Contrastive instruction evaluation method for better instruction tuning.
\newblock \emph{arXiv:2309.02301}, 2023.

\bibitem[Hyeon-Woo et~al.(2025)Hyeon-Woo, Ye-Bin, Choi, Hyun, and Oh]{hyeon-woo2025vlms}
Nam Hyeon-Woo, Moon Ye-Bin, Wonseok Choi, Lee Hyun, and Tae-Hyun Oh.
\newblock {VLM}{\textquoteright}s eye examination: Instruct and inspect visual competency of vision language models.
\newblock \emph{Transactions on Machine Learning Research}, 2025.

\bibitem[Hyun et~al.(2024)Hyun, Sung-Bin, Han, Yu, and Oh]{hyun-etal-2024-smile}
Lee Hyun, Kim Sung-Bin, Seungju Han, Youngjae Yu, and Tae-Hyun Oh.
\newblock {SMILE}: Multimodal dataset for understanding laughter in video with language models.
\newblock In \emph{NAACL Findings}, pages 1149--1167, 2024.

\bibitem[Jung et~al.(2025)Jung, Jang, and Chung]{jung2025avcd}
Chaeyoung Jung, Youngjoon Jang, and Joon~Son Chung.
\newblock {AVCD}: Mitigating hallucinations in audio-visual large language models through contrastive decoding.
\newblock In \emph{NeurIPS}, 2025.

\bibitem[Jung-Mok et~al.(2025)Jung-Mok, Hyeon-Woo, Ye-Bin, Nam, and Oh]{jung-mok2025automated}
Lee Jung-Mok, Nam Hyeon-Woo, Moon Ye-Bin, Junhyun Nam, and Tae-Hyun Oh.
\newblock Automated model discovery via multi-modal \& multi-step pipeline.
\newblock In \emph{NeurIPS}, 2025.

\bibitem[Kim et~al.(2025)Kim, Piergiovanni, Mallya, and Angelova]{Kim_2025_CVPR}
Dahun Kim, AJ Piergiovanni, Ganesh Mallya, and Anelia Angelova.
\newblock Videocomp: Advancing fine-grained compositional and temporal alignment in video-text models.
\newblock In \emph{CVPR}, pages 29060--29070, 2025.

\bibitem[Kim et~al.(2026)Kim, Seong-Eun, Jung-Mok, and Oh]{Kim_2026_WACV}
Kyeong~Seon Kim, Baek Seong-Eun, Lee Jung-Mok, and Tae-Hyun Oh.
\newblock m{EOL}: Training-free instruction-guided multimodal embedder for vector graphics and image retrieval.
\newblock In \emph{WACV}, pages 1191--1200, 2026.

\bibitem[Leng et~al.(2025)Leng, Xing, Cheng, Zhou, Zhang, Li, Zhao, Lu, Miao, and Bing]{leng2025the}
Sicong Leng, Yun Xing, Zesen Cheng, Yang Zhou, Hang Zhang, Xin Li, Deli Zhao, Shijian Lu, Chunyan Miao, and Lidong Bing.
\newblock The curse of multi-modalities: Evaluating hallucinations of large multimodal models across language, visual, and audio.
\newblock In \emph{NeurIPS Datasets and Benchmarks Track}, 2025.

\bibitem[Li et~al.(2025)Li, Zhang, Guo, Zhang, Li, Zhang, Zhang, Zhang, Li, Liu, and Li]{li2025llavaonevision}
Bo Li, Yuanhan Zhang, Dong Guo, Renrui Zhang, Feng Li, Hao Zhang, Kaichen Zhang, Peiyuan Zhang, Yanwei Li, Ziwei Liu, and Chunyuan Li.
\newblock {LL}a{VA}-onevision: Easy visual task transfer.
\newblock \emph{Transactions on Machine Learning Research}, 2025.

\bibitem[Li et~al.(2023{\natexlab{a}})Li, Wong, Zhang, Usuyama, Liu, Yang, Naumann, Poon, and Gao]{li2023llavamed}
Chunyuan Li, Cliff Wong, Sheng Zhang, Naoto Usuyama, Haotian Liu, Jianwei Yang, Tristan Naumann, Hoifung Poon, and Jianfeng Gao.
\newblock {LL}a{VA}-med: Training a large language-and-vision assistant for biomedicine in one day.
\newblock In \emph{NeurIPS Datasets and Benchmarks Track}, 2023{\natexlab{a}}.

\bibitem[Li et~al.(2023{\natexlab{b}})Li, Du, Zhou, Wang, Zhao, and Wen]{li-etal-2023-evaluating}
Yifan Li, Yifan Du, Kun Zhou, Jinpeng Wang, Xin Zhao, and Ji-Rong Wen.
\newblock Evaluating object hallucination in large vision-language models.
\newblock In \emph{EMNLP}, pages 292--305, 2023{\natexlab{b}}.

\bibitem[Lin et~al.(2024)Lin, Ye, Zhu, Cui, Ning, Jin, and Yuan]{lin-etal-2024-video}
Bin Lin, Yang Ye, Bin Zhu, Jiaxi Cui, Munan Ning, Peng Jin, and Li Yuan.
\newblock Video-{LL}a{VA}: Learning united visual representation by alignment before projection.
\newblock In \emph{EMNLP}, pages 5971--5984, 2024.

\bibitem[Liu et~al.(2023)Liu, Li, Wu, and Lee]{liu2023visual}
Haotian Liu, Chunyuan Li, Qingyang Wu, and Yong~Jae Lee.
\newblock Visual instruction tuning.
\newblock In \emph{NeurIPS}, 2023.

\bibitem[Liu et~al.(2024)Liu, Li, Li, and Lee]{Liu_2024_CVPR}
Haotian Liu, Chunyuan Li, Yuheng Li, and Yong~Jae Lee.
\newblock Improved baselines with visual instruction tuning.
\newblock In \emph{CVPR}, pages 26296--26306, 2024.

\bibitem[openAI(2025)]{gpt5}
openAI.
\newblock Introducing gpt-5.
\newblock \url{https://openai.com/index/introducing-gpt-5/}, 2025.

\bibitem[Radevski et~al.(2025)Radevski, Popordanoska, Blaschko, and Tuytelaars]{radevski2025dave}
Gorjan Radevski, Teodora Popordanoska, Matthew~B. Blaschko, and Tinne Tuytelaars.
\newblock {DAVE}: Diagnostic benchmark for audio visual evaluation.
\newblock In \emph{NeurIPS Datasets and Benchmarks Track}, 2025.

\bibitem[Radford et~al.(2021)Radford, Kim, Hallacy, Ramesh, Goh, Agarwal, Sastry, Askell, Mishkin, Clark, Krueger, and Sutskever]{pmlr-v139-radford21a}
Alec Radford, Jong~Wook Kim, Chris Hallacy, Aditya Ramesh, Gabriel Goh, Sandhini Agarwal, Girish Sastry, Amanda Askell, Pamela Mishkin, Jack Clark, Gretchen Krueger, and Ilya Sutskever.
\newblock Learning transferable visual models from natural language supervision.
\newblock In \emph{ICML}, pages 8748--8763, 2021.

\bibitem[Radford et~al.(2023)Radford, Kim, Xu, Brockman, Mcleavey, and Sutskever]{pmlr-v202-radford23a}
Alec Radford, Jong~Wook Kim, Tao Xu, Greg Brockman, Christine Mcleavey, and Ilya Sutskever.
\newblock Robust speech recognition via large-scale weak supervision.
\newblock In \emph{ICML}, pages 28492--28518, 2023.

\bibitem[Sakshi et~al.(2025)Sakshi, Tyagi, Kumar, Seth, Selvakumar, Nieto, Duraiswami, Ghosh, and Manocha]{sakshi2025mmau}
S Sakshi, Utkarsh Tyagi, Sonal Kumar, Ashish Seth, Ramaneswaran Selvakumar, Oriol Nieto, Ramani Duraiswami, Sreyan Ghosh, and Dinesh Manocha.
\newblock {MMAU}: A massive multi-task audio understanding and reasoning benchmark.
\newblock In \emph{ICLR}, 2025.

\bibitem[Schiappa et~al.(2022)Schiappa, Vyas, Palangi, Rawat, and Vineet]{schiappa2022robustness}
Madeline~Chantry Schiappa, Shruti Vyas, Hamid Palangi, Yogesh~S Rawat, and Vibhav Vineet.
\newblock Robustness analysis of video-language models against visual and language perturbations.
\newblock In \emph{NeurIPS Datasets and Benchmarks Track}, 2022.

\bibitem[Sun et~al.(2024)Sun, Yu, Tang, Chen, Tan, Li, Lu, Ma, Wang, and Zhang]{pmlr-v235-sun24l}
Guangzhi Sun, Wenyi Yu, Changli Tang, Xianzhao Chen, Tian Tan, Wei Li, Lu Lu, Zejun Ma, Yuxuan Wang, and Chao Zhang.
\newblock video-{SALMONN}: Speech-enhanced audio-visual large language models.
\newblock In \emph{ICML}, pages 47198--47217, 2024.

\bibitem[Sung-Bin et~al.(2025)Sung-Bin, Hyun-Bin, Lee, Senocak, Chung, and Oh]{sung-bin2025avhbench}
Kim Sung-Bin, Oh Hyun-Bin, JungMok Lee, Arda Senocak, Joon~Son Chung, and Tae-Hyun Oh.
\newblock {AVHB}ench: A cross-modal hallucination benchmark for audio-visual large language models.
\newblock In \emph{ICLR}, 2025.

\bibitem[Tang et~al.(2025)Tang, Li, Yang, Zhuang, Sun, Li, Ma, and Zhang]{tang2025video}
Changli Tang, Yixuan Li, Yudong Yang, Jimin Zhuang, Guangzhi Sun, Wei Li, Zejun Ma, and Chao Zhang.
\newblock video-salmonn 2: Caption-enhanced audio-visual large language models.
\newblock \emph{arXiv preprint arXiv:2506.15220}, 2025.

\bibitem[Team(2024)]{silero_vad2024}
Silero Team.
\newblock Silero vad: pre-trained enterprise-grade voice activity detector (vad), number detector and language classifier.
\newblock \url{https://github.com/snakers4/silero-vad}, 2024.

\bibitem[Touvron et~al.(2023)Touvron, Lavril, Izacard, Martinet, Lachaux, Lacroix, Rozière, Goyal, Hambro, Azhar, Rodriguez, Joulin, Grave, and Lample]{touvron2023llama}
Hugo Touvron, Thibaut Lavril, Gautier Izacard, Xavier Martinet, Marie-Anne Lachaux, Timothée Lacroix, Baptiste Rozière, Naman Goyal, Eric Hambro, Faisal Azhar, Aurelien Rodriguez, Armand Joulin, Edouard Grave, and Guillaume Lample.
\newblock Llama: Open and efficient foundation language models.
\newblock \emph{arXiv preprint arXiv:2302.13971}, 2023.

\bibitem[Xu et~al.(2025{\natexlab{a}})Xu, Guo, He, Hu, He, Bai, Chen, Wang, Fan, Dang, Zhang, Wang, Chu, and Lin]{xu2025qwen25omnitechnicalreport}
Jin Xu, Zhifang Guo, Jinzheng He, Hangrui Hu, Ting He, Shuai Bai, Keqin Chen, Jialin Wang, Yang Fan, Kai Dang, Bin Zhang, Xiong Wang, Yunfei Chu, and Junyang Lin.
\newblock Qwen2.5-omni technical report.
\newblock \emph{arXiv preprint arXiv:2503.20215}, 2025{\natexlab{a}}.

\bibitem[Xu et~al.(2025{\natexlab{b}})Xu, Guo, Hu, Chu, Wang, He, Wang, Shi, He, Zhu, Lv, Wang, Guo, Wang, Ma, Zhang, Zhang, Hao, Guo, Yang, Zhang, Ma, Wei, Bai, Chen, Liu, Wang, Yang, Liu, Ren, Zheng, Men, Zhou, Yu, Yang, Yu, Zhou, and Lin]{xu2025qwen3}
Jin Xu, Zhifang Guo, Hangrui Hu, Yunfei Chu, Xiong Wang, Jinzheng He, Yuxuan Wang, Xian Shi, Ting He, Xinfa Zhu, Yuanjun Lv, Yongqi Wang, Dake Guo, He Wang, Linhan Ma, Pei Zhang, Xinyu Zhang, Hongkun Hao, Zishan Guo, Baosong Yang, Bin Zhang, Ziyang Ma, Xipin Wei, Shuai Bai, Keqin Chen, Xuejing Liu, Peng Wang, Mingkun Yang, Dayiheng Liu, Xingzhang Ren, Bo Zheng, Rui Men, Fan Zhou, Bowen Yu, Jianxin Yang, Le Yu, Jingren Zhou, and Junyang Lin.
\newblock Qwen3-omni technical report.
\newblock \emph{arXiv preprint arXiv:2509.17765}, 2025{\natexlab{b}}.

\bibitem[Yang et~al.(2025)Yang, Li, Yang, Zhang, Hui, Zheng, Yu, Gao, Huang, Lv, Zheng, Liu, Zhou, Huang, Hu, Ge, Wei, Lin, Tang, Yang, Tu, Zhang, Yang, Yang, Zhou, Zhou, Lin, Dang, Bao, Yang, Yu, Deng, Li, Xue, Li, Zhang, Wang, Zhu, Men, Gao, Liu, Luo, Li, Tang, Yin, Ren, Wang, Zhang, Ren, Fan, Su, Zhang, Zhang, Wan, Liu, Wang, Cui, Zhang, Zhou, and Qiu]{yang2025qwen3technicalreport}
An Yang, Anfeng Li, Baosong Yang, Beichen Zhang, Binyuan Hui, Bo Zheng, Bowen Yu, Chang Gao, Chengen Huang, Chenxu Lv, Chujie Zheng, Dayiheng Liu, Fan Zhou, Fei Huang, Feng Hu, Hao Ge, Haoran Wei, Huan Lin, Jialong Tang, Jian Yang, Jianhong Tu, Jianwei Zhang, Jianxin Yang, Jiaxi Yang, Jing Zhou, Jingren Zhou, Junyang Lin, Kai Dang, Keqin Bao, Kexin Yang, Le Yu, Lianghao Deng, Mei Li, Mingfeng Xue, Mingze Li, Pei Zhang, Peng Wang, Qin Zhu, Rui Men, Ruize Gao, Shixuan Liu, Shuang Luo, Tianhao Li, Tianyi Tang, Wenbiao Yin, Xingzhang Ren, Xinyu Wang, Xinyu Zhang, Xuancheng Ren, Yang Fan, Yang Su, Yichang Zhang, Yinger Zhang, Yu Wan, Yuqiong Liu, Zekun Wang, Zeyu Cui, Zhenru Zhang, Zhipeng Zhou, and Zihan Qiu.
\newblock Qwen3 technical report.
\newblock \emph{arXiv preprint arXiv:2505.09388}, 2025.

\bibitem[Ye-Bin et~al.(2024)Ye-Bin, Hyeon-Woo, Choi, and Oh]{yebin2024beaf}
Moon Ye-Bin, Nam Hyeon-Woo, Wonseok Choi, and Tae-Hyun Oh.
\newblock {BEAF}: Observing before-after changes to evaluate hallucination in vision-language models.
\newblock In \emph{ECCV}, pages 232--248, 2024.

\bibitem[Ye-Bin et~al.(2025)Ye-Bin, Miles, Oh, Elezi, and Deng]{yebin2025retouchllmtrainingfreecodebasedimage}
Moon Ye-Bin, Roy Miles, Tae-Hyun Oh, Ismail Elezi, and Jiankang Deng.
\newblock Retouch{LLM}: Training-free code-based image retouching with vision language models.
\newblock \emph{arXiv preprint arXiv: 2510.08054}, 2025.

\bibitem[Zhang et~al.(2023)Zhang, Li, and Bing]{zhang-etal-2023-video}
Hang Zhang, Xin Li, and Lidong Bing.
\newblock Video-{LL}a{MA}: An instruction-tuned audio-visual language model for video understanding.
\newblock In \emph{EMNLP System Demonstrations}, pages 543--553, 2023.

\bibitem[Zhou et~al.(2018)Zhou, Xu, and Corso]{zhou2018towards}
Luowei Zhou, Chenliang Xu, and Jason~J. Corso.
\newblock Towards automatic learning of procedures from web instructional videos.
\newblock In \emph{AAAI}, 2018.

\bibitem[Zhou et~al.(2024)Zhou, Arnab, Buch, Yan, Myers, Xiong, Nagrani, and Schmid]{Zhou_2024_CVPR}
Xingyi Zhou, Anurag Arnab, Shyamal Buch, Shen Yan, Austin Myers, Xuehan Xiong, Arsha Nagrani, and Cordelia Schmid.
\newblock Streaming dense video captioning.
\newblock In \emph{CVPR}, pages 18243--18252, 2024.

\bibitem[Zhu et~al.(2025)Zhu, LIU, Zhang, Wang, Yangxue, Chen, Wang, Luo, and Zhang]{zhu2025alleviating}
Chenyu Zhu, YEFENG LIU, Hao Zhang, Aowen Wang, Yangxue, Guanhua Chen, Longyue Wang, Weihua Luo, and Kaifu Zhang.
\newblock Alleviating hallucinations in large language models through multi-model contrastive decoding and dynamic hallucination detection.
\newblock In \emph{NeurIPS}, 2025.

\bibitem[Zhu et~al.(2024)Zhu, Chen, Shen, Li, and Elhoseiny]{zhu2024minigpt}
Deyao Zhu, Jun Chen, Xiaoqian Shen, Xiang Li, and Mohamed Elhoseiny.
\newblock Mini{GPT}-4: Enhancing vision-language understanding with advanced large language models.
\newblock In \emph{ICLR}, 2024.

\bibitem[Zou et~al.(2025)Zou, Wang, Yan, Lyu, Zheng, Huang, Chen, Jiang, Liu, Tang, and Hu]{zou2025look}
Xin Zou, Yizhou Wang, Yibo Yan, Yuanhuiyi Lyu, Kening Zheng, Sirui Huang, Junkai Chen, Peijie Jiang, Jia Liu, Chang Tang, and Xuming Hu.
\newblock Look twice before you answer: Memory-space visual retracing for hallucination mitigation in multimodal large language models.
\newblock In \emph{ICML}, 2025.

\end{thebibliography}
}

\clearpage
\setcounter{page}{1}
\maketitlesupplementary

\setcounter{section}{0}
\renewcommand{\thesection}{\Alph{section}}
\hypersetup{linkcolor=black}

\begin{figure*}[!htbp]
    \centering
    \includegraphics[width=0.85\linewidth]{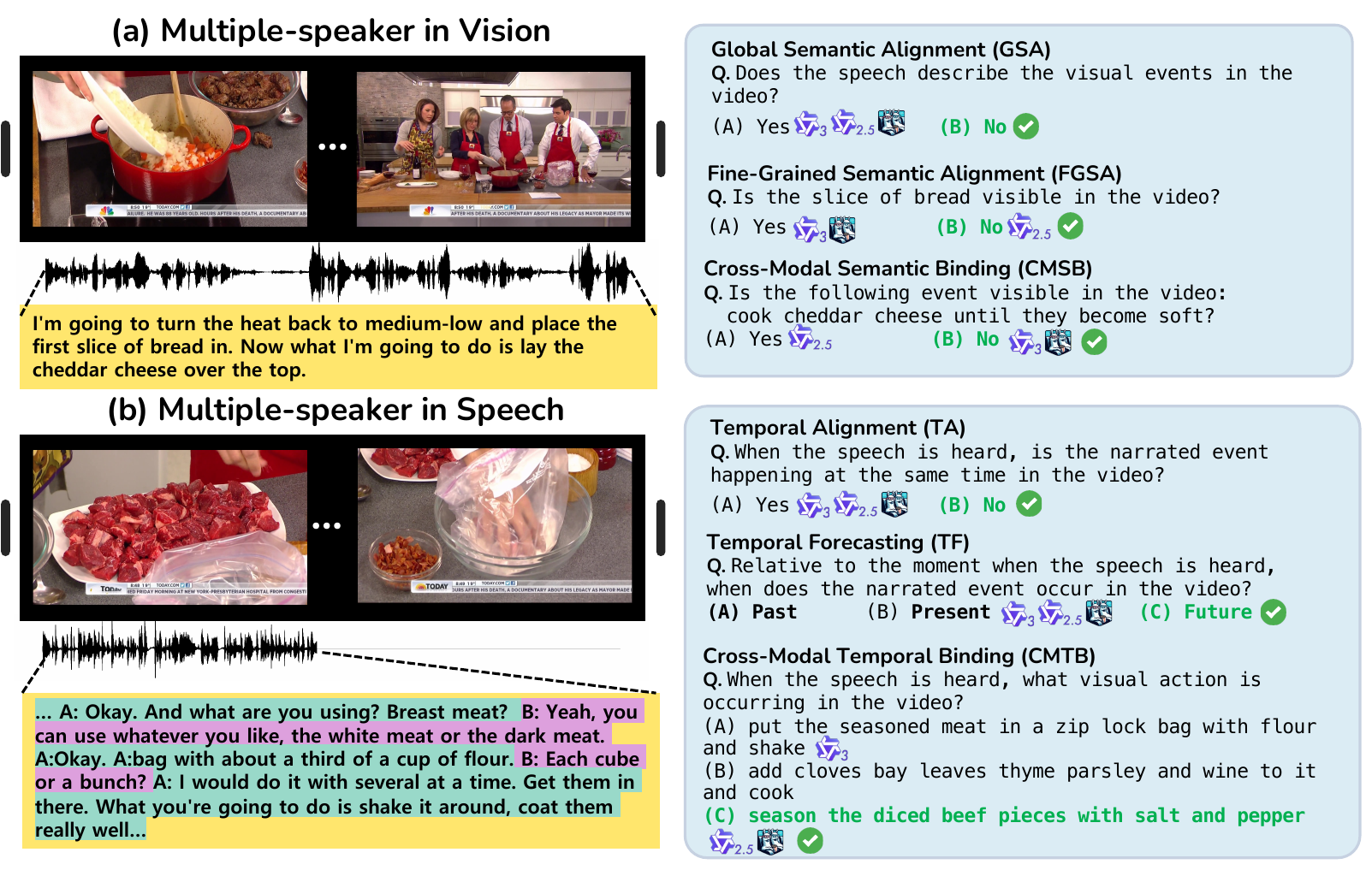}
    \caption{
    \textbf{Speaker variability in  \oursName{}.} We show predictions of Qwen3-Omni~\cite{xu2025qwen3} \qwenthree,  Qwen2.5-Omni~\cite{xu2025qwen25omnitechnicalreport} \qwentwo, and  Video-LLaMA 2~\cite{cheng2024videollama}\llama in diverse speaker settings.
    }
    \label{fig:speaker}
\end{figure*}

\vspace{-10pt}
\begin{figure*}[t]
    \centering
    \includegraphics[width=0.8\linewidth]{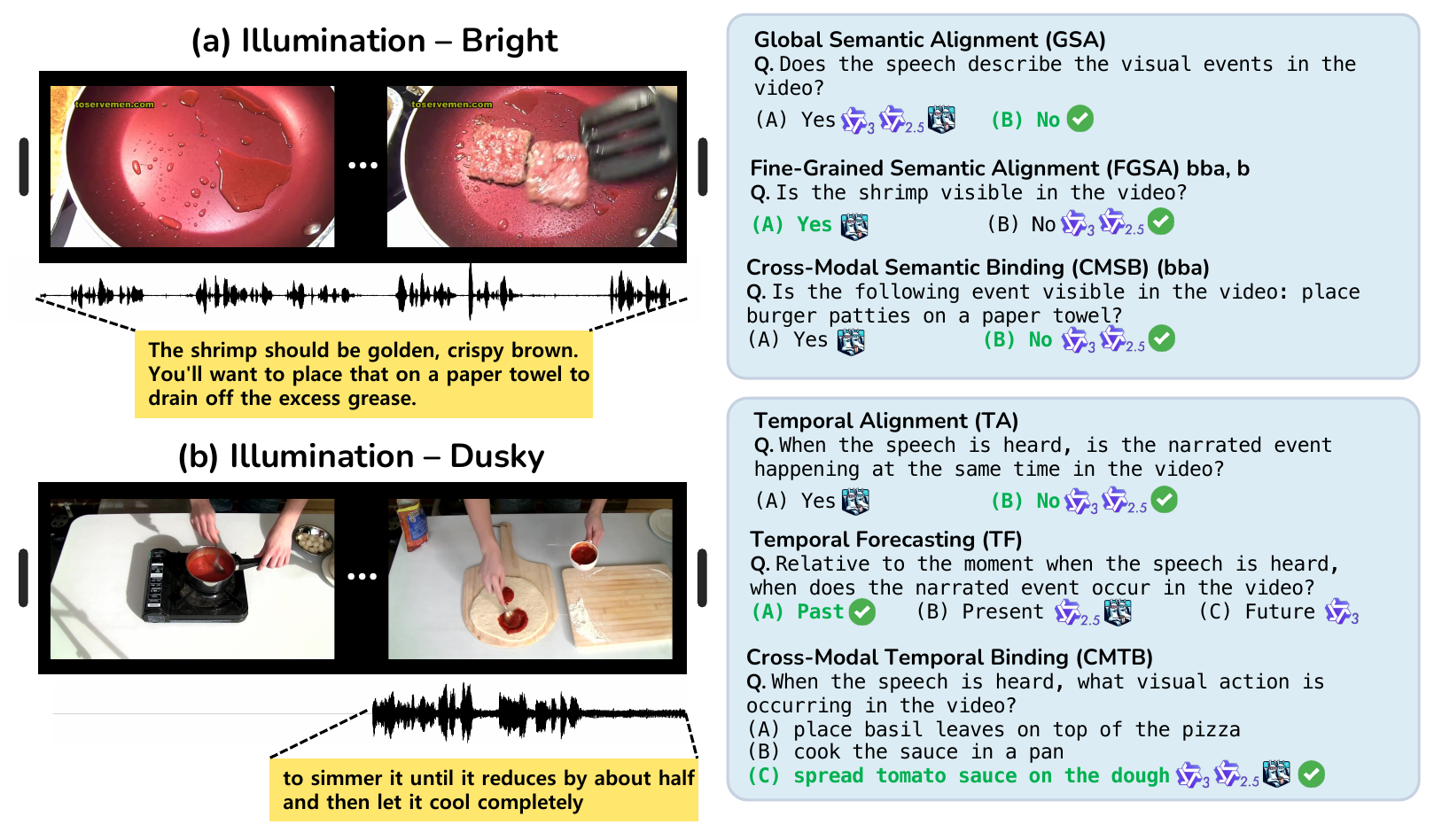}
    \caption{
      \textbf{Illumination variability in  \oursName{}.}  We show predictions of Qwen3-Omni~\cite{xu2025qwen3} \qwenthree,  Qwen2.5-Omni~\cite{xu2025qwen25omnitechnicalreport} \qwentwo, and  Video-LLaMA 2~\cite{cheng2024videollama}\llama in diverse illumination settings.
}
    \label{fig:lighting}
\end{figure*}

\begin{figure*}[ht]
    \centering
    \includegraphics[width=0.9\linewidth]{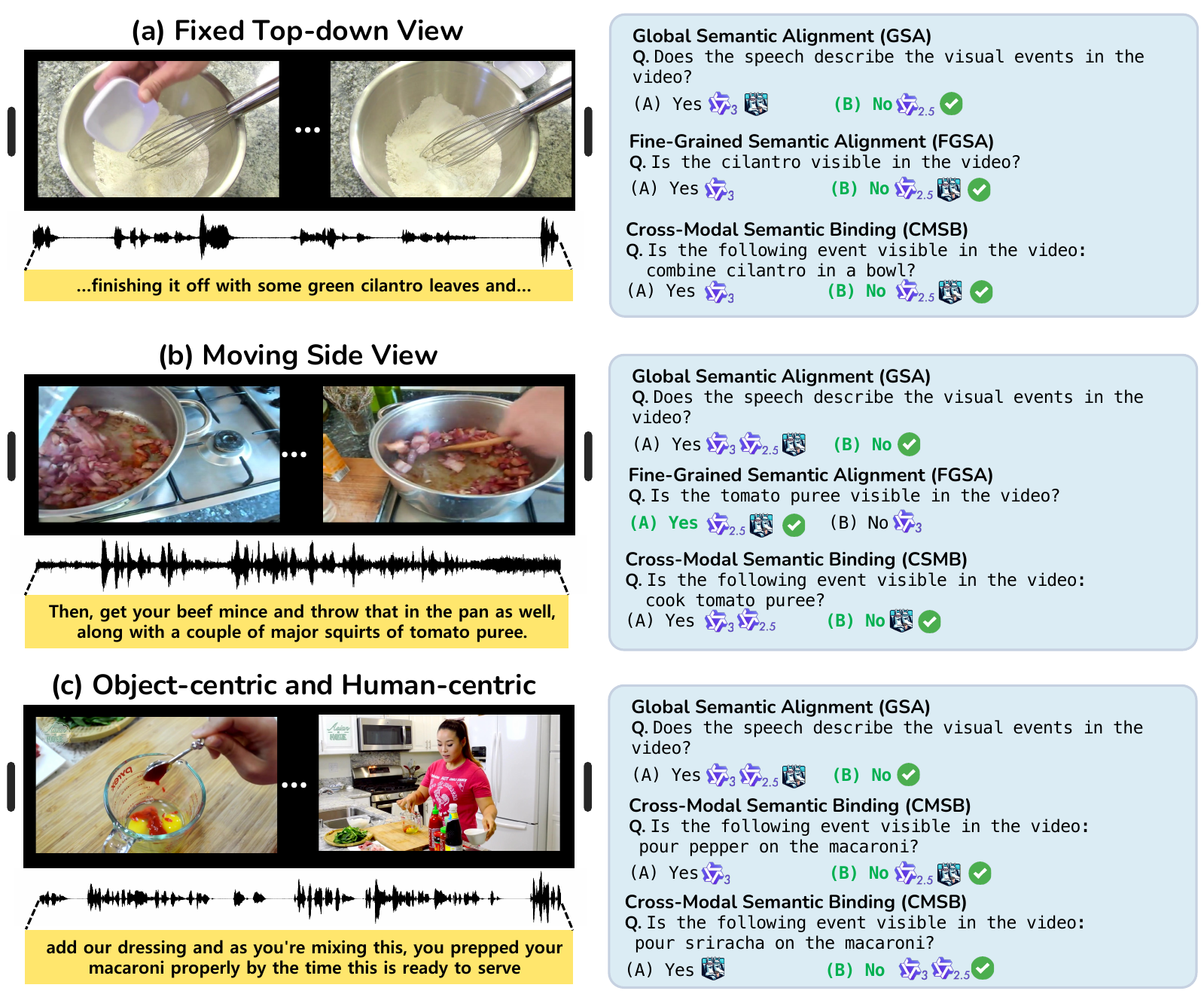}
    \vspace{-10pt} 
    \caption{
         \textbf{Camera viewpoint variability in  \oursName{}.}   We show predictions of Qwen3-Omni~\cite{xu2025qwen3} \qwenthree,  Qwen2.5-Omni~\cite{xu2025qwen25omnitechnicalreport} \qwentwo, and  Video-LLaMA 2~\cite{cheng2024videollama}\llama in diverse camera viewpoint settings.
    }
    \label{fig:cameraview}
\end{figure*}

\section{Additional Qualitative Results}

In this section, we visualize video–question pairs from \oursName{}, together with predictions from the latest audio–visual large language models (LLMs). Our \oursName{} shows diversity in both task design and video scenarios, providing a comprehensive benchmark for evaluating speech-vision hallucination in audio-visual LLMs. 

\begin{itemize}

    \item \textbf{Speaker variability}. \oursName{}  includes videos where multiple people appear visually in the scene (Figure~\ref{fig:speaker}(a)) or only present in the audio (Figure~\ref{fig:speaker}(b)).  
    
    \item \textbf{Illumination variability}. Our dataset includes videos under diverse lighting conditions,  such as bright environments (Figure~\ref{fig:lighting}(a)) or dusky settings (Figure~\ref{fig:lighting}(b)).

        \item \textbf{Camera viewpoint diversity}. \oursName{} includes videos with different viewpoints, such as fixed top-down views (Figure~\ref{fig:cameraview}(a)), moving side-view (Figure~\ref{fig:cameraview}(b)), and transitions between object-centric and human-centric views (Figure~\ref{fig:cameraview}(c)).

\end{itemize}
 
\section{GPT Prompts for Dataset Creation}

During dataset creation, we adopt GPT~\cite{gpt5} models to extract the objects and actions for task formulation. To find objects and actions that uniquely appear in video or speech, we prompt the following instruction based on the caption of the video and speech. We find that GPT could return good results in most cases, and we perform a human verification based on GPT results.

\begin{tcolorbox}[listing style]
Given caption1 and caption2, extract action and objects. Action refers to verbs, e.g., put in the bowl. \\
First, extract all actions and objects from each caption. Second, find unique actions and objects that only appear in one caption but not the other, termed as $cap1\_only_{act}$, $cap1\_only_{obj}$, $cap2\_only_{act}$, $cap2\_only_{obj}$, respectively.

Following is one example. Output in the json format. 

Input:

caption1: put the eggs in the bowl
caption2: cut the eggs and cucumber

Output:

\{
  $cap1\_only_{act}$: ["put in the bowl"], \\
  $cap1\_only_{obj}$: [], \\
  $cap2\_only_{act}$: ["cut"], \\
  $cap\_only_{obj}$: ["cucumber"], \\
\}
\end{tcolorbox}

\section{Dataset Quality Control}
To ensure the dataset quality, we design diverse strategies to filter out unqualified samples, including rule-based strategy and GPT-based strategy. For example, we filter out short videos and speeches to avoid the case that the events are not clearly shown. We also filter out samples if the speech only includes general information, e.g., ``this looks great'', instead of describing the events.  After dataset creation, we perform a final human verification to ensure the dataset quality. The high performance of Gemini-2.5 Pro also indicates the high quality of our dataset, while highlighting the severe hallucinations of open-source models.

\end{document}